# A ROBUST CRYPTOGRAPHIC SYSTEM USING NEIGHBORHOOD-GENERATED KEYS

Samuel King Opoku

*Computer Science Department, Kumasi Polytechnic, Ghana*
*Email: samuel.k.opoku@gmail.com*

**Abstract:** The ability to hide information from unauthorized individuals has been a prevalent issue over the years. Countless algorithms such as DES, AES and SHA have been developed. These algorithms depend on varying key length and key management strategies to encrypt and decrypt messages. The size of the encrypted message is so large that it therefore consumes and wastes valuable storage space when implemented in organizations that store and handle large volumes of small data. The ability to share the generated keys securely also poses a problem. This paper proposes a robust cryptographic algorithm which generates its keys from the surroundings and already-designed coding schemes. The proposed system conserves storage space and processing power. The algorithm is implemented and tested using PHP and MySQL DBMS.

## I. INTRODUCTION

Cryptography has been the technique for preventing unwanted disclosure of the content of information through encryption. Encryption is therefore the method of transforming information into an illegible format. Cryptography involves the use of keys in either symmetric or asymmetric way to encrypt messages [1]. In symmetric cryptography, the receiver and sender of an encrypted message have the same key [1], [2] to decipher the encrypted message. The commonly known algorithms that use symmetric cryptography are Data Encryption Standards (DES) and Advanced Encryption Standard (AES). The problem associated with symmetric cryptography is the ability to share the key securely [2].

Asymmetric cryptography, on the other hand, requires the generation of two keys – public and private keys – for data encryption. The public key may be distributed freely and is used for encrypting data. The private key is given to the intended recipient of the encrypted data who can decrypt the data securely [3]. Traditional asymmetric cryptography widely and effectively used in the Internet relies on a Public Key Infrastructure (PKI). The success of PKI depends on the availability and security of a Certificate Authority (CA) – a central control point that everyone trusts [4]. To ensure maximum security, private and public keys generated by PKI are periodically refreshed as per application by the Trent [3], [4]. Trent regenerates the keys and informs the parties repeatedly. Each party's public key is distributed to the entire parties and the private key needs to be sent to the party individually via a secure channel. Key distribution in asymmetry cryptography is much easier. Authentication and non-repudiation are available in asymmetric key distribution. Compromising a private key of a user does not reveal messages encrypted for other users in the group. The involvement of Trent in generating keys however affects the privacy issue of the individuals [3]. Asymmetric cryptography is generally computational expensive [4].

Asymmetric cryptography is used in Digital Signature Standard (DSS), Password authenticated key agreement technique and with key distribution algorithms [1], [3], [5]. Commonly known algorithms that use asymmetric cryptography are Elliptic Curve Cryptosystem (ECC), Rivest-Shamir-Adleman (RSA) [6]. Identity-Based Cryptography (IBC) has been the method developed to eliminate the requirement of certificate distribution [7] - [9]. IBC brings free pair-wise keys without any interaction between nodes requiring lower processing power, storage space and communication bandwidth. It however requires signature scheme [7]. Protocols based on certificateless public key cryptography (CL-PKC) are preferred since they do not need certificates to guarantee the authenticity of public keys and does not suffer from key escrow of identity-based cryptography [3], [5]

Data can be encrypted in three ways [10], [11]:
- Stream Cipher which works by encrypting single bits or bytes of information one a time. The technique uses symmetric cryptography methods in its process
- Block Cipher which is an encryption technique that breaks down and encrypts data into individual blocks. The size of a block is typically 64bits [12]. The DES and AES have used block cipher design in their applications [2].
- Using hash algorithm is another encryption technique. It uses mathematical algorithm to

convert large amounts of data into smaller, compressed size. This technique is used in "One way encryption" [13] where data is not decrypted for viewing. It provides data integrity and authentication but no confidentiality. The general hashing algorithms uses 128-bit hash value, 160-bit hash or 256-bit hash value. 128-bit hash values are produced by Message Digest (MD) algorithms such as MD2, MD4 and MD5. 160-bit hash values are produced by Secure Hash Algorithms (SHA) such as SHA-1. Various versions of SHA's with varying hash values have been evolved. These include SHA-256, SHA-384 and SHA-512 which respectively produce 256-bit, 384-bit and 512-bit hash values. This method generate large values for small data and is therefore not appropriate for encrypting large number of small data that need to be stored and retrieved in a database

The figure below summarizes the functionalities of the above algorithms.

| Algorithm Type | Encryption | Digital Signature | Hashing Function | Key Distribution |
|---|---|---|---|---|
| **Asymmetric Key Algorithms** | | | | |
| RSA | X | X | | X |
| ECC | X | X | | X |
| Diffie-Hellman | | | | X |
| El Gamal | X | X | | X |
| DSA | | X | | |
| LUC | X | X | | X |
| Knapsack | X | X | | X |
| **Symmetric Key Algorithms** | | | | |
| DES | X | | | |
| 3DES | X | | | |
| Blowfish | X | | | |
| IDEA | X | | | |
| RC4 | X | | | |
| SAFER | X | | | |
| **Hashing Algorithms** | | | | |
| Ronald Rivest family of hashing functions: MD2, MD4, and MD5 | | | X | |
| SHA | | | X | |

*Figure 1: Functionalities of the Various Cryptographic Algorithms*

In spite of the above algorithms, encrypted messages are usually subjected to attacks. Active attackers modify system files, masquerade and change messages with the aim to discover the key used in the encryption process. Commonly used attacks are [14]-[18]:

- Cipher-Only (also known as Known-Cipher) Attacks: Attacker has ciphertext of several messages that have been encrypted using the same encryption algorithm. The attacker then aims at deducing plaintext or even better, the key from the ciphertext.

- Known-Plaintext Attacks: Attacker has the plaintext and ciphertext of one or more messages. The attacker tries to discover the key used to encrypt one of the messages and use the key to decipher and read the other messages. A typical variation is the Linear Cryptanalysis in which the attacker evaluates the input and output values for each Substitution-box (S-box) [19]. With S-box, each message is being encrypted with the same key [20]. The attacker evaluates the probability of input values ending up in a specific combination and assigns these probability values to different keys until one shows a continual pattern of having the highest probability

- Chosen-Plaintext Attacks: Attacker has the plaintext but can choose randomly the plaintext to be encrypted to see the corresponding ciphertext. This gives the attacker, more power and possibly a deeper understanding of the way the encryption process works so that they gather more information about the key being used. Another form of this attack is the Differential Cryptanalysis (DC). In DC, the attacker takes two messages of plaintext and follows the changes that took place to the blocks as they go through different S-boxes. The differences identified in the resulting ciphertext values are used to map probability values to different possible key values. DC has been used to break several cryptosystems such as DES and Triple-DES [21], RC4 [22] and AES [23].

- Chosen-Ciphertext Attacks: Attacker chooses a ciphertext to be decrypted under unknown key. The goal is to figure out the key. This is a harder attack to carry out compared to the previously mentioned attacks. The attacker needs to have control of the system that contains the cryptosystem

- Side-Channel Attacks: Attacker measures power consumption, radiation emissions and the time it takes for certain types of data to process. With this information, an attacker can work backwards by reverse-engineering the process to uncover an encryption key or sensitive data. A power attack, an attack to review the amount of heat released, has been successful in uncovering confidential information from smart cards. RSA private keys were uncovered in 1995 by measuring the relative time cryptographic operations took

- Statistical Attacks: Attacker identifies statistical weaknesses in algorithm design for exploitation. For instance, a Random Number Generation (RNG) may be biased. If keys are taken directly

- from the output of the RNG, then the distribution of keys would also be biased. The statistical knowledge about the bias could be used to reduce the search time for the keys

- Replay Attack: Attacker captures some type of data (usually authenticated information) and resubmits it with the hope of fooling the receiving device into thinking that it is legitimate information. Time stamps and sequence numbers are countermeasures to replay attacks.

The strength of a cryptographic mechanism depends on the length of the key and the key management strategy used [11]. The length of the key is directly proportional to the effectiveness of the algorithm and the number of bits in the encrypted message. Thus storage space are usually consumed and wasted when these algorithms are implemented in organizations that store and use large volumes of small data. This paper proposes a robust cryptographic algorithm which generates its keys from the surroundings and coding schemes. The proposed system conserves storage space and processing power. The algorithm is implemented and tested using Hypertext Preprocessor (PHP) and MySQL Database Management System (DBMS).

## II. CRYPTOGRAPHIC PARAMETERS AND KEYS GENERATION

This section describes the various elements required to generate the keys for the cryptographic system. It also describes how these keys are obtained from the database environment.

### A. Obtaining Cryptographic Parameters

The cryptographic system designed in this work is based on the following:

- The database table which contains the amount to be encrypted has Date, Time and the Cashier's Identification (CashierID) as part of its fields. These fields may constitute the primary fields of the table

- The Date should be in the format dd-mm-yy or dd/mm/yy. The separator of the date elements is inconsequential to the efficiency of the system. The elements (dd, mm, yy) are employed such that $\forall dd,\ 01 \leq dd \leq 31$; $\forall mm,\ 01 \leq mm \leq 12$ and $\forall yy,\ 00 \leq yy \leq 99$.
  Designers can choose yyyy such that $\forall yyyy,\ 0000 \leq yyyy \leq 9999$ instead of yy.

- The Time should be in the format HH:MM:SS and that $\forall HH,\ 00 \leq HH \leq 23$; $\forall MM,\ 01 \leq MM \leq 59$ and $\forall SS,\ 00 \leq SS \leq 59$

- The CashierID consists of uppercase alphanumeric characters only. There is no such special symbol as +, =, * and -. Thus CashierID only contains English uppercase letters and digits. Software designers are also allowed to include lowercase letters and then assign coding scheme values to them.

- The amount to be encrypted should be in the format $\#^+.\#\#$, where $\# \in Z^+$ and $0 \leq \# \leq 9$ and the Kleene star, +, represents one or more occurrences. Thus the least amount is 0.00

### B. Key Generation

The keys are generated from the fields in the database table. The various elements (dd, mm, yy or yyyy, HH, MM and SS) are manipulated such that the value obtained for any given expression is greater than zero. Basically, there are two keys generated. These are called $K_1$ and $K_2$. $K_1$ is obtained from date and time elements whereas $K_2$ is obtained from CashierID.

To obtain the value of $K_1$ from date and time, designers are allowed to create their own formulas but they should keep in mind that their formulas should satisfy this condition, $K_1 > 0$. A typical generation of $K_1$ which is used throughout this work is given as:

$$K_1 = (dd * mm + SS) * (HH + MM + yy).$$

With this definition of $K_1$, it implies that $\forall K_1$, $K_1 > 0$ and $1 \leq K_1 \leq 78011$. $K_1 = 1$ when the date is given as 01/01/00 (that is 1st January, 2000) and the time of entry is 00:00:00 (that is 12:00:00 am). However, $K_1 = 78011$ when the date is 31/12/99 (that is 31st December, 1999) and the time is 23:59:59. Since date and time do not remain constant, it implies that the same amount encrypted at different periods will have different $K_1$ values and hence different cipher texts.

The key, $K_2$, is obtained from the individual characters which constitute CashierID so that:

$$K_2 = \sum (CashierID\ Coded\ Characters).$$

Each character in CashierID is encoded by assigning them numeric values. The flexibility of the system also gives the designer the choice of values to be assigned to the various characters. However, to improve the efficiency of the system, designers are encouraged to assign at least two-digit value (that is a value greater than or equal to ten) to each character. The table below shows a sample of values assigned to each encoded character.

*Table 1: Sample Coding Scheme for $K_2$ Generation*

| CODING SCHEME FOR GENERATING $K_2$ | | | |
|---|---|---|---|
| A → 21 | J → 30 | S → 39 | 1 → 48 |
| B → 22 | K → 31 | T → 40 | 2 → 49 |

| C → 23 | L → 32 | U → 41 | 3 → 50 |
|---|---|---|---|
| D → 24 | M → 33 | V → 42 | 4 → 51 |
| E → 25 | N → 34 | W → 43 | 5 → 52 |
| F → 26 | O → 35 | X → 44 | 6 → 53 |
| G → 27 | P → 36 | Y → 45 | 7 → 54 |
| H → 28 | Q → 37 | Z → 46 | 8 → 55 |
| I → 29 | R → 38 | 0 → 47 | 9 → 56 |

### III. ENCRYPTION ALGORITHM

Given that any amount entered to be stored in a database is formatted as #⁺.##, the amount can be divided into two parts. These are the integer part and the fractional part. The integer part, $I_p$, is a number such that $I_p \geq 0$ whereas the fractional part, $F_p$, is also a number which satisfies $00 \leq F_p \leq 99$. Each part of the amount is encrypted differently with different procedure to ensure that ciphertext generated is relatively provable.

*A. Integer Part Encryption*

The integer part is encrypted using four simple but effective steps. Given the keys $K_1$ and $K_2$ and the amount to be encrypted which is denoted as Amt, the following steps are applied:

Step 1: Let $\emptyset$ be the size of the data type (like integer, float and double) of the variable (called Cipher1) used to store the value of the first step of the encryption process, then

$$\text{Cipher1} = n^p * K_1 * Amt \pm m^q * K_1$$

where $n, p, m\ and\ q$ are constants, $n^p$ and $m^q$ represent different coefficients of $K_1$ which are used to enhance the effectiveness of the algorithm such that:

- $\forall p,\ p \in Z^+ \geq 0$ and
  $$p < abs\left(Integer\left(\frac{\log \emptyset - \log K_1}{\log n}\right) - 1\right)$$
  where abs denote absolute value

- $\forall q,\ q \in Z^+ \geq 0$ and
  $$q < abs\left(Integer\left(\frac{\log \emptyset - \log K_1}{\log n}\right) - 1\right)$$

- $\forall m, n;\ m, n \in Z^+ \geq 1$.

The size of Cipher1 depends on the size of its data type. Thus for large values, designers are encouraged to use double data type instead of integer or even float

Step 2: Cipher2 = Cipher1 + $K_2$

Step 3: Count the number of characters in Cipher2. If the total number of characters is not even then pad Cipher2 with a zero. This is done by using the algorithm blow:

*/* converts Cipher2 to string using*
*% (a symbol for arithmetic modulo) and*
*+ (a symbol for concatenation) */*
**String str_Cipher2 = Cipher2 + "";**
**if (Length(str_Cipher2) % 2 != 0){**
   **str_Cipher2 = "0" + str_Cipher2;**
**}**

Step 4: Convert str_Cipher2 to Cipher3 using the coding scheme shown in the table below which is obtained by grouping str_Cipher2 characters into twos.

*Table 2: Sample Coding Scheme for Obtaining Cipher3*

| CODING SCHEME FOR CONVERTING str_Cipher2 TO Cipher3 | | | |
|---|---|---|---|
| 00 → + | 25 → h | 50 → U | 75 → ¬ |
| 01 → - | 26 → I | 51 → u | 76 → ` |
| 02 → * | 27 → i | 52 → V | 77 → 1 |
| 03 → ? | 28 → J | 53 → v | 78 → 2 |
| 04 → ! | 29 → j | 54 → W | 79 → 3 |
| 05 → , | 30 → K | 55 → w | 80 → 4 |
| 06 → [ | 31 → k | 56 → X | 81 → 5 |
| 07 → ( | 32 → L | 57 → x | 82 → 6 |
| 08 → ] | 33 → l | 58 → Y | 83 → 7 |
| 09 → ) | 34 → M | 59 → y | 84 → 8 |
| 10 → A | 35 → m | 60 → Z | 85 → 9 |
| 11 → a | 36 → N | 61 → z | 86 → 0 |
| 12 → B | 37 → n | 62 → . | 87 → ; |
| 13 → b | 38 → O | 63 → ^ | 88 → space |
| 14 → C | 39 → o | 64 → < | 89 → / |
| 15 → c | 40 → P | 65 → > | 90 → \ |
| 16 → D | 41 → p | 66 → % | 91 → \| |
| 17 → d | 42 → Q | 67 → = | 92 → { |
| 18 → E | 43 → q | 68 → $ | 93 → } |
| 19 → e | 44 → R | 69 → £ | 94 → " |
| 20 → F | 45 → r | 70 → _ | 95 → ' |
| 21 → f | 46 → S | 71 → # | 96 → € |
| 22 → G | 47 → s | 72 → & | 97 → α |
| 23 → g | 48 → T | 73 → ~ | 98 → β |
| 24 → H | 49 → t | 74 → @ | 99 → γ |

This step is employed to conserve storage space and also to discourage naive editing. The coding scheme ensures that instead of storing N characters (where N is the length of str_Cipher2), $N/2$ characters are stored. Users of the system who are attackers are discouraged from editing str_Cipher2 elements which are numbers only with the view that the numbers correspond to the correct amount. Encoding individual digits in str_Cipher2 without grouping them gives ten (10) ways to determine the digits that constitute str_Cipher2. However, grouping the digits into twos gives hundred (100) ways to determine any paired number which relatively increases the efficiency of the system. If the digits in str_Cipher2 are rotated or rearranged, then using individual digits without grouping them, there are $10N$ ways (where $N$ is the length of the str_Cipher2 or the total number of characters in str_Cipher2) to determine the digits in Cipher2 from Cipher3. The complexity of the algorithm in terms of decrypting time is $O(n)$.

However, when the digits in str_Cipher2 are grouped into twos, there are $100N(N/2)$ ways of determining the digits in Cipher2 from Cipher3. The complexity of the algorithm in terms of decrypting time is $O(n^2)$. The primary challenge of grouping the digits in str_Cipher2 into twos is the inclusion of non-English characters. The number of characters for a Standard English keyboard is ninety-seven (97) and thus there are three (3) other character groups that need to be handled. In this work, Greek letters were chosen and this required additional system configuration to handle Greek letters

*B. Fractional Part Encryption*

The fractional part is encrypted using coding scheme in Table 2. The keys $K_1$ and $K_2$ are not used in encrypting the fractional part of the amount. The fractional part of the amount is encrypted differently to frustrate attackers. Thus, the encoded fractional part forms Cipher4 and the resulting Ciphertext obtained from encrypting the integer part and the fractional part of the amount is given by:

$$\text{Ciphertext} = \text{Cipher4Cipher3}$$

*C. Demonstration of Encryption Algorithm*

To illustrate the encryption algorithm, consider an amount of GHS 50.00 paid on 21st May, 2012 at 15:50:08 received by a cashier with CasherID CE840716, using Table 1 and Table 2, the keys are computed as:

For $K_1$:
  Using yy format of the year, it implies dd=21, mm=5, yy=12, HH=15, MM=50 and SS = 8.
  Thus $K_1$ is computed as:
  $K_1 = (dd * mm + SS) * (HH + MM + yy)$
  $= (21 * 5 + 8) * (15 + 50 + 12)$
  $= 113 * 77$
  $= 8701$

For $K_2$:
  Using the coding scheme in Table 1, it implies C → 23, E → 25, 8 → 55, 4 → 51, 0 → 47, 7 → 54, 1 → 48 and 6 → 53. Thus $K_2$ is computed as
  $K_2 = \sum(CE840716)$
  $= (23 + 25 + 55 + 51 + 47 + 54 + 48 + 53)$
  $= 356$

Let $p = 0$ and $q = 0$ then $n^p = 1\ and\ m^q = 1$. Given that Amt = 50,
  Cipher1 = $K_1$ * Amt + $K_1$
  $= 8701 * 50 + 8701$
  $= 443751$

  Cipher2 = Cipher1 + $K_2$
  $= 443751 + 356$
  $= 444107$.

The length of Cipher2 is six which is even. Hence padding is ignored

Grouping Cipher2 into twos without rearranging them, we have 44, 41 and 07. Thus Cipher3 is obtained as:
  Cipher3 = Encode (Grouped Cipher2)
  = Encode (44, 41, 07)

using coding scheme in Table 2 so that 44→R, 41→p and 07→(
  = Rp(

The fractional part, 00, is encoded using coding scheme in Table 2 to get Cipher4. Thus
  Cipher4 = Encode (Fractional Part)
  = Encode (00)
  = +

Finally, the encrypted amount (called Ciphertext) which is stored in the database is obtained as:
  Ciphertext = Cipher4Cipher3
  = +Rp(

## IV. DECRYPTION ALGORITHM

The algorithm for decrypting a Ciphertext requires seven (7) steps which also involve data integrity checking. The description is used to decrypt +Rp( which was encrypted in section III above given that $p = 0,$ and $q = 0$ then $n^p = 1\ and\ m^q = 1$.

Step 1: Split the Ciphertext into two groups, mainly, $T_1$ and $T_2$ with $T_1$ containing only the first character of the Ciphertext and $T_2$ containing the remaining characters of the Ciphertext. The algorithm below demonstrates how splitting is accomplished in Java:
  $T_1$ = Ciphertext.substring (0, 1);
  $T_2$ = Ciphertext.substring (1);

Step 2: Decode $T_1$ by reversing the coded scheme in Table 2. For example, let $T_1$ = +. The value that corresponds to $T_1$ = + is 00. Thus Plain(+) = 00

Step 3: Decode $T_2$ using the reverse coded scheme in Table 2. Thus using Rp( as an example, R→44, p→41 and (→07. Thus Plaintext1($T_2$) = 444107

Step 4: Compute $K_2$ from the CashierID value of the CashierID field using coded scheme in Table 1. Deduct $K_2$ from Plaintext1($T_2$). Using the example above with $K_2$ = 356. It implies that

Plaintext2($T_2$) = 444107 – 356
= 443751.

Step 5: Compute $K_1$ using date and time values from the date and time fields respectively in the database. Use $K_1$ to find the Plaintext of the integer part as:

Plaintext(integer part) = (plaintext2($T_2$) – $K_1$) / $K_1$.

Using the example above such that $K_1$ = 8701, it implies that
Plaintext(integer part) = (443751 - 8701) / 8701
= 50

Step 6: The division part of the computation of Plaintext(integer part) which is shown as:
*Plaintext(integer part) = (plaintext2($T_2$) – $K_1$) / $K_1$*
is used to check data integrity. If the division results in a real value instead of integer, then the value in the database has been changed and an alarm needs to be raised. Otherwise go to step 7.

Step 7: This is the final stage where the fractional part and the integer part are put together.
Thus
Amount = Plaintext(integer).Plain($T_1$).

Hence using the above example, the amount is calculated as 50.00

## V. SYSTEM IMPLEMENTATION AND ALGORITHM TESTING

The system is implemented and tested via the development of an accounting system. The accounting system is used to collect school fees and also manages the financial system of the institution. The introduction of the Greek letters requires system setting modification. In creating a database using MySQL Database Management System, the language setting needs to be changed. The figure below demonstrates how to set the language of the database so as to accept Greek letters. The name of the database is smsystem.

```
create database if not exists `smsystem`
    DEFAULT CHARACTER SET greek COLLATE greek_general_ci;
```
*Figure 2: Database Creation*

A typical table is also created such that the language it accepts is also altered to accept Greek letters as well. The figure below demonstrates how this can be done.

```
/*Table structure for table 'payment'*/

DROP TABLE IF EXISTS 'payment';

CREATE TABLE 'payment' (
    'STDID' varchar(15) NOT NULL,
    'DATE' varchar(15) NOT NULL,
    'TIME' varchar(15) NOT NULL,
    'TERM' varchar(2) default NULL,
    'ACYEAR' varchar(5) default NULL,
    'RECEIPTNUMBER' varchar(30) default NULL,
    'MODE' varchar(10) default NULL,
    'AMOUNT' varchar(10) default NULL,
    PRIMARY KEY ('STDID', 'DATE', 'TIME'),
    CONSTRAINT 'payment_ibfk_1' FOREIGN KEY ('STDID')
        REFERENCES 'particulars'('STDID')
)ENGINE=InnoDB DEFAULT CHARACTER SET
    greek COLLATE greek_general_ci;
```
*Figure 3: Table Creation*

Working with web design tools to create web pages be client side like HyperText Markup Language (HTML) or server side scripting language like Hypertext Preprocessor (PHP) requires language modification as well. The figure below illustrates the language setting of a web page.

```
<html>
    <head>
        <meta http-equiv="Content-Type"
            content="text/html" charset="ISO-8859-7">
    </head>
    <body>
        <? php
            //your code to query the database
        ?>
        <!-- other codes of the program -->
    </body>
</html>
```
*Figure 4: Web Page Language Setting*

The system is designed to use $K_1$ and $K_2$ formulas that are described under section II of this paper. The system is implemented such that:
- There are fifteen different coding schemes from which $K_2$ can be obtained.
- There are fifteen different coding schemes for obtaining Cipher3 from str_Cipher2.
- Cipher1 is obtained from $K_1$ using the formula:
  $n^p * K_1 * Amt \pm m^q * K_1$ such that
  $\forall\, n, m;\, n \neq m$ and $m, n \in \{1, 2, 3, 4, \ldots, 50\}$

and $n^p \neq m^q = 1$. Also $\forall\, p, q;\ p, q \in \{0, 1, 2, 3, 4\}$ and $p \neq q = 0$

The system is implemented so that either n, m, p, q or the coding schemes used are changed periodically to increase the effectiveness of the system. Whenever the parameters are changed, all the encrypted amount in the system's database are updated based on the new values of the cryptographic parameters. A table called smsystem_specifications is created which has only one field, namely smsystem_settings. The field contains the string in the order p,q,m,n,z,w where p, q, m and n are the respective values of the constants p, q, m and n used as the current cryptographic parameters. z is the number of the coding scheme used for obtaining $K_2$. The coding schemes are numbered 1 to 15. w is the number of the coding scheme used for converting str_Cipher2 to Cipher3. The coding schemes are numbered 1 to 15. These coding schemes are hidden from users and they are hard-coded. The algorithm below illustrates how the system determines the next cryptographic parameters:

```
01 Start
02   Load Prev_P, Prev_Q, Prev_M,
03   Load Prev_N, Prev_Z, Prev_W
04
05   //computing old parameters
06   Compute Prev_n^p and Prev_m^q
07
08   //computing new parameter
09   do
10     new_P = Math.round (Math.random() * 4)
11   while (new_P < 0)
12
13   do
14     new_Q = Math.round (Math.random() * 4)
15   while (new_Q < 0)
16
17   if (new_P == 0 and new_Q == 0)
18     Go to line 09
19
20   do
21     new_N = Math.round (Math.random() * 50)
22   while (new_N < 1)
23
24   do
25     new_M = Math.round (Math.random() * 50)
26   while (new_M < 1)
27
28   Compute new_n^p and new_m^q
29   if (new_n^p == 1 and new_m^q == 1 )
30     Go to line 20
31
32   do
33     new_Z = Math.round (Math.random() * 15)
34   while (new_Z < 1)
35
36   do
37     new_W = Math.round (Math.random() * 15)
38   while (new_W < 1)
39
40   //compare values
41   if (new_P == Prev_P and new_Q == Prev_Q and
42     new_N == Prev_N and new_M  Prev_M and
43     new_Z == Prev_Z and new_W == Prev_W)
44     Go to line 09
45   else
46     for all encrypted amount
47       decipher each amount using Prev_n^p, Prev_m^q,
48         Prev_Z and Prev_W
49       encrypt each amount again using new_n^p,
50         new_m^q, new_Z and new_W
51     end for
52   end if
53 Stop
```

## VI. ANALYSIS OF THE IMPLEMENTED SYSTEM

### A. Strengths of the System

The strengths of the system are analyzed as follow:

- Encryption uses multiplication and addition which prevent the occurrence of fraction in its computation. However, decryption uses division. The resulting value of the division, *(plaintext2($T_2$) $\pm$ $m^q*K_1$) / ($n^p*K_1$)*, should always be an integer. The system has a simple way of checking data integrity. If a value is changed, the outcome of its decryption shall not be integer but real

- Unlike the widely known algorithms which are made up of at least sixteen characters occupying device storage space that can be used for storing other important information, this design has a minimum of three ciphertext characters and a maximum of $\left(\frac{\text{length of Amount } + 5}{2}\right) \in \mathbb{Z}^+$ ciphertext characters given that CashierID contains three to seventeen characters and $p = q = 0$.

- Keys are generated from the surroundings, that is other fields in the same database table. The field values form part of the record. Without generating different keys which are sometimes stored alongside the ciphertext conserves space and controls processing power

- The manipulative or computational part of the algorithm can be change anytime without affecting the efficiency and effectiveness of the algorithm. For instance, the format for computing $K_1$ can be changed so as the coding schemes

### B. Limitations of the System

The system also has limitations and they are discussed as follow:

- The CashierID should be in uppercase characters only in this implementation. Although the character coding scheme can be modified to handle lowercase character coding as well, the processing power for searching characters and the memory for storing character coding will be unnecessarily consumed

- To accommodate Greek letters, the Database Management System and the front-end programming language require extra settings which usually consume designer's time

- Careful manipulation of ciphertext can result in integer value division which will adversely compromise the integrity of the amount stored. This problem is solved by periodically changing the key format especially $K_1$. Frequent application of this mechanism will considerably slows down the system.

## VII. CONCLUSION

A new concept for cryptosystem has been proposed based on surrounding parameters. The system gives several advantages from a technical viewpoint: flexibility, robustness and operational efficiency. The system serves as a basis for implementing cryptosystems that generate their keys from the surroundings. This eliminates problems that emanate from sharing cryptographic keys between end points. Future works will focus on using intelligent systems such as Artificial Neural Networks to encrypt messages using keys generated from the surroundings.